\documentclass[aoas, reqno]{imsart}

\usepackage{amsthm,amsmath,natbib}
\RequirePackage[colorlinks,citecolor=blue,urlcolor=blue]{hyperref}
\usepackage{amsfonts, graphicx, mathtools, float, color} 
\graphicspath{{fig/}}

\startlocaldefs
\usepackage{latent_social_movement}
\endlocaldefs

\begin{document}

\begin{frontmatter}

  \title{Dynamic social networks based on movement}
\runtitle{Dynamic social networks based on movement}

\author{\fnms{Henry R.}\snm{Scharf}\thanksref{m1,t1}\corref{}
\ead[label=e1]{henry.scharf@colostate.edu}},
\author{\fnms{Mevin B.}
  \snm{Hooten}\thanksref{m4,m1}\ead[label=e2]{mevin.hooten@colostate.edu}},
\author{\fnms{Bailey K.}
  \snm{Fosdick}\thanksref{m1}\ead[label=e3]{bailey.fosdick@colostate.edu}},
\author{\fnms{Devin S.} \snm{Johnson}\thanksref{m2}\ead[label=e4]{}},
\author{\fnms{Josh M.} \snm{London}\thanksref{m2}\ead[label=e5]{}}, \and
\author{\fnms{John W.} \snm{Durban}\thanksref{m3}\ead[label=e6]{}}

\thankstext{t1}{Corresponding author} 

\runauthor{H. Scharf et al.}

\affiliation{Colorado State University\thanksmark{m1}, U. S. Geological Survey,
  Colorado Cooperative Fish and Wildlife Research Unit\thanksmark{m4}, NOAA
  Alaska Fisheries Science Center\thanksmark{m2}, and NOAA Southwest
  Fisheries Science Center\thanksmark{m3}}


\begin{abstract}
  Network modeling techniques provide a means for quantifying social structure
  in populations of individuals. Data used to define social connectivity are
  often expensive to collect and based on case-specific, \textit{ad hoc}
  criteria. Moreover, in applications involving animal social networks,
  collection of these data is often opportunistic and can be
  invasive. Frequently, the social network of interest for a given population is
  closely related to the way individuals move. Thus telemetry data, which are
  minimally-invasive and relatively inexpensive to collect, present an
  alternative source of information. We develop a framework for using telemetry
  data to infer social relationships among animals. To achieve this, we propose
  a Bayesian hierarchical model with an underlying dynamic social network
  controlling movement of individuals via two mechanisms: an attractive effect,
  and an aligning effect. We demonstrate the model and its ability to accurately
  identify complex social behavior in simulation, and apply our model to
  telemetry data arising from killer whales. Using auxiliary information about
  the study population, we investigate model validity and find the inferred
  dynamic social network is consistent with killer whale ecology and expert
  knowledge.
\end{abstract}


\end{frontmatter}

\section{Introduction}
Dynamic social networks are an important topic of study among ecologists for a
variety of species and ecological processes
(\citealt{pinter-wollman_dynamics_2013, krause_social_2007,
  croft_exploring_2008, wey_social_2008, sih_social_2009}). Social networks can
help explain a myriad of behavioral activities in a population, including the
characteristics of animal movement. Therefore, it is common to define social
networks based on directly observable behavior such as the duration of time
animals spend in close proximity to one another (e.g., African elephants,
\textit{Loxodonta africana}, \citealt{goldenberg_controlling_2014}), discrete
counts of interactions (e.g., yellow (\textit{Papio cynocephalus}) and anubis
baboons (\textit{Papio anubis}) \citealt{franz_self-organizing_2015}), or
discrete counts of close encounters (e.g., barn swallows (\textit{Hirundo
  rustica erythrogaster}) \citealt{levin_performance_2015}). Challenges for
researchers interested in studying animal social networks include expensive data
collection procedures, and potential biases due to opportunistic observation.

Killer whales (\textit{Orcinus orca}), like many marine mammals, are complex and
highly social creatures (\citealt{pitman_cooperative_2012, Parsons2009,
  williams_killer_2006, baird_social_2000}). To better understand the behavior
of killer whales, we seek to characterize their social
relationships. Unfortunately, direct observation of killer whale interactions is
challenging; it is not uncommon for individuals to travel 50km a day and to
range over thousands of kilometers in a season (\citealt{durban_antarctic_2012,
  andrews_satellite_2008}). Furthermore, observation of killer whales at close
proximity has been found to significantly influence their movement behavior
(\citealt{williams_behavioural_2002}), which could directly affect measurements
of social connectivity. In contrast, satellite tracking tags have been used to
gather movement data for killer whales over several months
(\citealt{durban_antarctic_2012, andrews_satellite_2008}), and there is little
evidence to suggest that tags alter behavior. Thus, a potential alternative to
costly personal observations are telemetry data, which contain rich movement
information at the individual level, and can be collected in remote areas at a
much lower cost.

The suite of models for animal telemetry data is vast and rapidly changing,
including both continuous- and discrete-time approaches (see
\citealt{mcclintock_when_2014} for a review). Yet there are only a few models
that explicitly account for interactions among individuals in the population
(e.g., \citealt{russell_dynamic_2015, langrock_modelling_2014,
  codling_copycat_2014, morales_building_2010}). Moreover, methods are lacking
that attempt to characterize pairwise connections between all members of the
population. We propose a model for movement that incorporates plausible
mechanistic effects on movement due to an underlying social network. Our model
allows us to infer the specific characteristics of interaction in a given
population and the underlying dynamic social network itself.

In our proposed discrete-time continuous-space model, we assume there exists an
underlying (latent) dynamic social network among the individuals in the
population. Conditional on the network characteristics and the positions of
animals in the previous time step, the expected positions of individuals at the
next time point are modeled jointly using a Gaussian Markov random field (GMRF)
(\citealt{besag_spatial_1974, besag_conditional_1995, rue_gaussian_2005}). The
model is temporally Markovian for both the animal positions and the social
network. In our model, the underlying social structure influences movement
through two channels: an attractive effect and an alignment effect. These
channels of interaction allow us to model a wide variety of behaviors, and they
have a precedent for use in the context of interaction behavior
(\citealt{lemasson_motion-guided_2013}). The connection between the underlying
social network and position is an example of a hidden Markov model (HMM). HMMs
represent a flexible class of hierarchical models popular in analyses of
wildlife data (see, for example, \citealt{langrock_flexible_2012}) in which an
observable process (in our case, position) is driven by an unobserved Markovian
process (the underlying social network).

We introduce the details of our proposed method in Section~\ref{sec:methods}. We
demonstrate and assess inference from the model with simulated data in
Section~\ref{sec:simulation}.  In Section~\ref{sec:killer_whales}, we analyze
data for seven killer whales tagged concurrently near the coast of the Antarctic
Peninsula. Within the tagged sample, there are three genetically distinct types
of killer whale (\citealt{pitman_three_2003, Morin2015}) characterized by their
size, coloration, and diet. The spatial distributions for each type overlap, and
while strong social interaction is typical within each type, there have been no
observed social associations among animals of different types. We demonstrate
that inferences from our method are consistent with this history of
observation. Furthermore, we find strong evidence for dynamic social connections
forming and dissolving within each type, but no indication of connections
between types. Finally, in Section \ref{sec:discussion}, we discuss potential
extensions for the model, including the incorporation of environmental
covariates and approaches for mediating the large computational demands for the
model when the study sample is large.

\section{Methods}
\label{sec:methods}
We propose new methodology based on a general hierarchical modeling framework
that accommodates measurement, process, and parameter uncertainty
(\citealt{berliner_hierarchical_1996}). We introduce the GMRF that describes
animal movement in Section \ref{sec:process_position} and describe our method
for modeling the dynamic social network in Section
\ref{sec:process_network}. Then in Section \ref{sec:data_crawl}, we detail how
we account for the fact that telemetry data are typically measured at
individual-specific, irregularly spaced times with error.

\subsection{Position Process}
\label{sec:process_position}
A GMRF is a description of a Gaussian random vector where conditional dependence
between elements is specified based on a neighborhood structure
(\citealt{rue_gaussian_2005}). For example, data occurring at regular intervals
in time, or on a lattice in space, are often modeled with GMRFs because natural
neighborhoods exist for each datum (e.g., the preceding measurement in time, or
the four closest spatial locations). Thus, GMRFs present a natural mathematical
structure for modeling trajectories of connected individuals, as they provide a
way to model dependence between connected or ``neighboring'' individuals.

We expect that social structure among individuals will influence their movement
with respect to one another. Let $\muvec_i(t)$ denote the position of individual
$i$ at time $t$. Assuming we know the population social structure (i.e., which
individuals are socially affiliated with which other individuals), we model the
movements of all individuals simultaneously using a GMRF involving two social
behavioral mechanisms: one related to attraction toward the mean position of
connected individuals, and the other related to alignment, or movement parallel
to the paths taken by connected individuals. Although our model is flexible
enough to capture attraction or repulsion, as well as alignment or
anti-alignment, in most cases, we expect to infer assortative relations whereby
individuals that are socially connected move ``together." For this reason, we
discuss movement of connected individuals as aligned and attractive.

Attraction and alignment mechanisms are critical features of the mean positions
of each individual at regular synchronous time steps. Models for locations on
regular intervals have been developed by several others, including
\citet{brillinger_elephant-seal_1998}, \citet{jonsen_robust_2005}, and
\citet{forester_statespace_2007}. We define the social relations in terms of a
dynamic binary network $\W(t)$ indexed at times $t = 1, \dots, T$, where entry
$w_{ij}(t) = 1$ indicates a connection between individuals $i$ and $j$ at time
$t$ and $w_{ij}(t) = 0$ indicates a lack thereof.

We specify a GMRF conditionally, from the perspective of a single individual at
a given time. The mean position of each individual $i$ at time $t$ conditioned
on all other individuals' positions at time $t$, denoted $\muvec_{-i}(t)$, and
all positions at the previous time, $\muvec(t - 1)$, follows a normal
distribution with mean

\begin{align}
  &\E \lp \muvec_i(t) | \muvec_{-i}(t), \muvec(t - 1), \W(t), \W(t - 1),
  \alpha, \beta, \sigma^2, c \rp \equiv \nonumber \\
  &\hspace{-.2in} \muvec_i(t - 1)
  + \underbrace{\beta \muvectil_i(t - 1)}_\text{attraction}
  + \underbrace{\sum_{j \neq i} \alpha \frac{w_{ij}(t)}{w^c_{i+}(t)} \lp
    \muvec_j(t) - (\muvec_j(t - 1) + \beta \muvectil_j(t - 1))
    \rp}_\text{alignment} \label{eqn:cond_mean}
\end{align}

and precision

\begin{align}
  &\text{Prec}\lp \muvec_i(t)|\W(t),
  \sigma^2, c \rp \equiv \sigma^{-2}w^c_{i+}(t) \I_2. \label{eqn:precision}
\end{align}

Focusing on \eqref{eqn:cond_mean}, we model the expected location of individual
$i$ as the sum of three terms: the individual's location in the previous time
period, $\muvec_i(t-1)$; an attraction term capturing the tendency for the
individual to move toward other individuals it is socially connected to; and an
alignment term accounting for groups of interconnected individuals moving in
the same general direction.

The term $\muvectil_i(t)$, in the attraction component of \eqref{eqn:cond_mean},
is a unit vector pointing from individual $i$'s position $\muvec_i(t)$ to the
mean position $\muvecbar_i(t)$ of all the individuals it is connected to in
$\W(t)$ (i.e., its ego-network):

\begin{align}
  \muvecbar_i(t) &\equiv
  \sum_{j \neq i}^n \frac{w_{ij}(t)}{w^c_{i+}(t)}\muvec_j(t) 
  \label{eqn:muvecbar} \\
  \muvectil_i(t) &\equiv
  \begin{cases}
    \frac{\muvecbar_i(t) - \muvec_i(t)}{||\muvecbar_i(t) - \muvec_i(t)||_2},
    \quad & \sum_{j \neq i}w_{ij}(t) > 0\\
    0, & \sum_{j \neq i}w_{ij}(t) = 0.
  \end{cases}
\end{align}

The parameter $\beta$ controls the strength of the attractive effect of a social
connection. On average, individual $i$ moves a distance $\beta$ in the direction
$\muvectil_i(t)$ during each time step.

In the above expression, $w^c_{i+}(t)$ is the size of individual $i$'s
ego-network at time $t$ if the individual has at least one connection (i.e.,
$w^c_{i+}(t) = \sum_{j \neq i}w_{ij}(t)$), and equal to a constant
$w^c_{i+}(t)=c>0$ otherwise. We require $c$ to be strictly positive so the
precision in \eqref{eqn:precision} is non-zero for unconnected individuals.

The alignment term in \eqref{eqn:cond_mean} quantifies the mean displacement in
position from $t-1$ to $t$ for only those individuals that individual $i$ is
socially connected to, and after accounting for attraction. Although the sum is
over all individuals $j$, the social network indicators $w_{ij}(t)$ eliminate
the effects of an individual's direction if it is not connected to individual
$i$. The parameter $\alpha$ controls the strength of the aligning effect, with 0
corresponding to no alignment, and $\alpha \rightarrow 1$ corresponding to
perfect alignment. The case $\alpha = 1$ corresponds to an intrinsic conditional
autoregressive model with an improper covariance matrix. However, we limit our
consideration to $\alpha < 1$, precluding this special case.

Finally, the expression for the precision in \eqref{eqn:precision} has the
property that individuals who are more socially connected (i.e., have larger
ego-networks $w_{i+}^c(t)$), have larger precision. The proportional
relationship between precision and $w_{i+}^c(t)$ is required for a valid GMRF,
and aligns with our intuition that, conditioned on the position of all other
individuals, the movement of an individual with few or no social connections is
more difficult to predict than one that experiences strong attraction and
alignment toward a large group of individuals. The parameter $c$ can be thought
of as the effective size of the ego-network for an unconnected individual with
regard to precision.

The specification of the model in \eqref{eqn:cond_mean} and
\eqref{eqn:precision} properly defines a GMRF where the elements of the
precision matrix at time $t$ are

\begin{align}
  Q_{ij}(t) &\equiv \begin{cases}
    -\alpha w_{ij}(t)\sigma^{-2}\I_2, \quad & j \neq i \\
    w^c_{i+}(t)\sigma^{-2}\I_2, & j = i.
  \end{cases}
\end{align}

Therefore, we can write the multivariate version of the model for $t = 2, \dots,
T$ as

\begin{align}
  \lb \muvec(t)|\muvec(t - 1), \thetavec \rb \equiv
  \N(\muvec(t - 1) + \beta \muvectil(t - 1), \; \Q(t)),& \label{eqn:mu_t}
\end{align}

where we have concatenated the model parameters $\lp \alpha, \beta, p_1, \phi,
\sigma^2, c, \W \rp$ into a single vector $\thetavec$ (note: $p_1$ and $\phi$
are parameters associated with the dynamic network $\W$ and are introduced in
Section \ref{sec:process_network}).

\begin{figure}[ht]
  \centering
  \includegraphics[width = \linewidth]{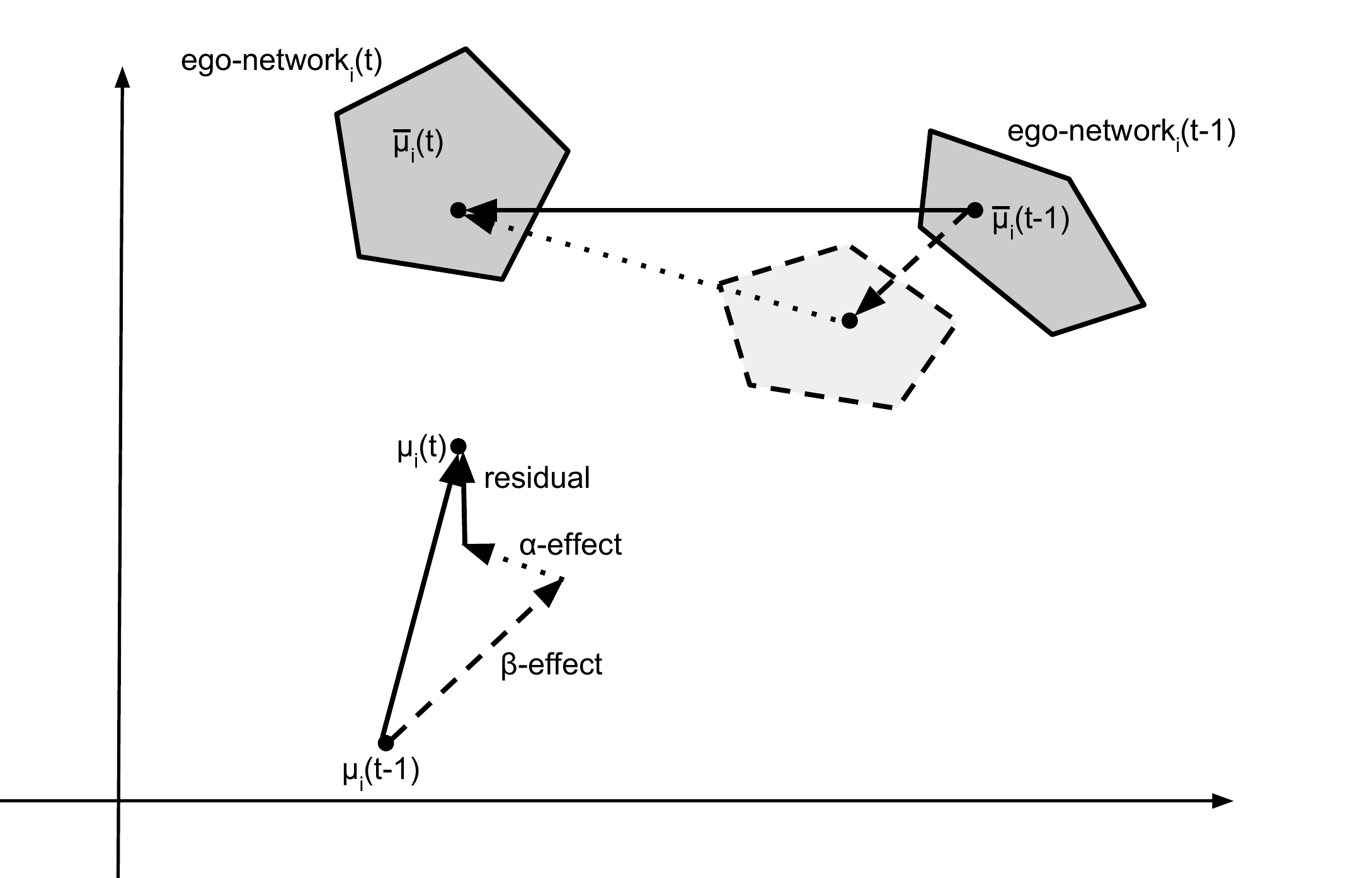}
  \caption{This schematic illustrates the two channels through which the dynamic
    social network influences movement. The dashed lines represent where the
    ego-network of individual $i$ would be expected to be at time $t$ under
    attraction alone, and the parallel dotted lines represent the alignment
    between individual $i$ and the average of the differences $\muvec_j(t) -
    (\muvec_j(t - 1) + \beta \muvectil_j(t - 1))$.}
  \label{fig:schematic}
\end{figure}

Notice that, for the joint distribution in \eqref{eqn:mu_t}, the attraction
effect remains in the mean structure because the attraction force for an
individual is toward the previous location of the individuals in the
ego-network. However, the alignment effects are accounted for in the precision
matrix because alignment is characterized by simultaneous movement of grouped
individuals in the same direction. Figure \ref{fig:schematic} shows the
alignment and attraction effects graphically.

The model for movement based on the normalized vector $\muvectil_i(t)$, instead
of $\muvecbar_i(t) - \muvec_i(t)$, reflects a mechanistic understanding that
attractive movement is often restricted by the distance an animal can reasonably
travel in a given time step. We assume the maximum distance an individual is
capable of moving during one time step to be approximately constant. Thus, when
the gap between an individual and the center of its ego-network is large
compared to its step size, an animal feeling an attractive pull will appear to
take several steps of similar length in that direction.

If we had used the difference $\muvecbar_i(t) - \muvec_i(t)$ instead of $\muvectil_i(t)$ in the attraction component of \eqref{eqn:cond_mean}, the attractive
pull an individual experienced when its ego-network was far away could be far
greater than the distance it was able to travel in a single time step. To see
this, note that the interpretation of $\beta$ in \eqref{eqn:mu_t} would change
to reflect the average proportion of the gap between an individual and the
center of its ego-network covered during each time step. A value of $\beta =
0.5$ would imply that an animal closes half the distance between itself and the
center of its ego-network, regardless of the size of that gap. In some cases,
the proportional gap coverage model may be more appropriate. In our application
with killer whales, it is reasonable for connections between animals to form
across relatively large gaps in space relative to the distance an animal might
be able to cover in a single time step. Thus, the former interpretation is the
most appropriate for our application.

In \eqref{eqn:cond_mean} and \eqref{eqn:muvecbar} we define the vector
$\muvecbar_i(t - 1)$ using the status of the social network at time $t -
1$. Another possibility is to define $\muvecbar_i(t - 1)$ using the social network at the current time $t$ as

\begin{align}
  \muvecbar_i(t - 1) &\equiv
  \sum_{j \neq i}^n \frac{w_{ij}(t)}{w^c_{i+}(t)}\muvec_j(t - 1).
  \label{eqn:ae_option}
\end{align}

In practice, the differences that arise in the estimated social network
depending on this modeling decision will only be noticeable near times when a
connection status changes (i.e., whenever $\mathbf{w}(t) \neq \mathbf{w}(t -
1)$). Hence, when the estimated social network is slowly varying, like the one
we observe in our application, we expect that these two definitions will result
in essentially identical inference. However, for applications when the frequency
of changes in social connections is high relative to the scale at which
telemetry observations are made, the impact of the decision of how to define
$\muvecbar$ may be more significant.

\subsection{Dynamic Social Network}
\label{sec:process_network}
We model the dynamic process that gives rise to $\W(t)$ as a collection of
pairwise independent Bernoulli random variables with a Markovian dependence in
time, where

\begin{align}
  w_{ij}(1) &\sim \Bern(p_1) \label{eqn:w1} \\
  w_{ij}(t)|w_{ij}(t - 1) &\sim
  \begin{cases}
    \Bern(p_{1|0}), \; &w_{ij}(t - 1) = 0\\
    \Bern(p_{1|1}), &w_{ij}(t - 1) = 1
  \end{cases} \qquad t = 2, \dots, T. \label{eqn:wt}
\end{align}

The parameter $p_1$ is the probability of a social connection between any two
individuals at time $t = 1$, $p_{1|0}$ is the conditional probability that a
pair of individuals who are not connected at $t - 1$ become connected at time
$t$, and $p_{1|1}$ is the conditional probability that a pair connected at time
$t - 1$ remain connected at time $t$. Thus, our model for $\muvec$ can be
thought of as a HMM, where the latent social network $\W$ takes on the role of
the hidden Markovian process. Though the model for the dynamic social network
could be used exactly as specified in \eqref{eqn:w1} and \eqref{eqn:wt}, we make
two refinements that reduce the number of parameters we are required to
estimate, and facilitate solicitation of priors.

First, we take advantage of the fact that, in many cases, it is reasonable to
assume the mean density of a study population's dynamic social network remains
constant in time. This is equivalent to requiring that the stationary
distribution of the Markov process governing the overall network density match
the expected density at time $t=1$. Recall that we model the conditional
distributions of the edges, $\mathbf{w}(t) | \mathbf{w}(t - 1)$ as independent
Bernoulli random variables. Thus, the expected density of the network at time
$t$ is equal to the marginal probability of an edge between any two vertices,
$\Pr(w_{ij}(t) = 1)$. The Markov process controlling network density is
therefore the same as the process for the sequence of social connections
$\mathbf{w}_{ij}$. Requiring that the initial density, $p_1$, match the
stationary distribution of the Markov process is equivalent to the condition

\begin{align}
  p_1 &= \frac{p_{1|0}}{p_{1|0} + 1 - p_{1|1}} \label{eqn:stationarity}.
\end{align}

Condition \eqref{eqn:stationarity} implicitly reduces the number of
parameters to be estimated from three to two.

The second refinement we make is a reparameterization that allows for more
intuitive interpretation of model parameters, and hence, facilitates the
solicitation of priors. We define a new variable, $\phi$, that controls the
temporal stability of the dynamic network via

\begin{align}
  p_{1|0} &\equiv (1 - \phi)p_1,   \label{eqn:reparameter}
\end{align}

which implies we can write $p_{1|1} = 1 - (1 - \phi)(1 - p_1)$.  As $\phi$
varies from 0 to 1, the social network transitions smoothly from complete
temporal independence, to complete temporal dependence (i.e., a static network
where no edges form or dissolve in time). This can be expressed mathematically
as $\Lim{\phi \rightarrow 0} p_{1|0} = \Lim{\phi \rightarrow 0} p_{1|1} = p_1$
and $\Lim{\phi \rightarrow 1} \lp 1 - p_{1|0} \rp = \Lim{\phi \rightarrow 1}
p_{1|1} = 1.$ Thus, $\phi$ can be thought of as a measure of the temporal range
of dependence in the network. Under the parameterization using $p_1$ and $\phi$,
researchers can construct priors for the network density and stability
independently of one another.

\subsection{Measurement Error and Time Alignment}
\label{sec:data_crawl}
Our model can be used to make inference about the posterior distribution of the
model parameters $\thetavec$ conditioned on the mean position process $\muvec$
(denoted $\lb \thetavec | \muvec \rb$). However, in practice, we are rarely able
to observe $\muvec$ directly. Rather, we observe noisy measurements of position
at asynchronous, irregularly occurring times, which we denote $\svec$, and the inference we wish to make
is for the posterior distribution conditioned on observed data, not $\muvec$. Let
$\svec_i(\tau_i)$ denote the observed position of individual $i$ at time
$\tau_i$, and $\lb \svec | \muvec\rb$ the joint density of all observed
locations conditioned on the unobserved processes $\muvec$. The top level of our
hierarchical model provides a connection between the locations $\muvec_i(t)$,
which occur at regular synchronized times, and the observations $\svec_i(t)$.

We could obtain the desired posterior distribution by evaluating the integral

\begin{align}
  \lb \thetavec | \svec \rb &= \int \lb \thetavec | \muvec, \svec \rb
  \lb \muvec | \svec \rb d\muvec, \label{eqn:posterior}
\end{align}

using Markov chain Monte Carlo (MCMC), provided we could sample from the
distribution $\lb \muvec | \svec \rb$. Unfortunately, because of the inherent
complexities in the irregular, asynchronous observation times and the high
dimensionality of the vector $\muvec$, sampling from this distribution becomes
computationally infeasible when a study population contains more than a few
individuals and a few dozen observation times per individual. We address this
issue by making use of a multiple imputation procedure employed by
\citet{hooten_agent-based_2010} and \citet{hanks_restricted_2015,
  hanks_velocity-based_2011}, paired with a continuous-time correlated random
walk model from \citet{johnson_continuous-time_2008}. Multiple imputation offers a computationally efficient way to account for asynchronous, noisy position measurements while still permitting us to use a discrete-time, step-aligned structure for movement informed by a dynamic social network. We outline the procedure briefly below, and refer the reader to \citet{hooten_agent-based_2010} and \citet{hanks_velocity-based_2011} for further details.  

The premise of the multiple imputation strategy assumes the existence of
a distribution that is very similar to $\lb \muvec | \svec \rb$ from which we
can sample paths easily. If we can define such a distribution, which we call
$\lb \muvec^* | \svec \rb$, then we can closely approximate the integral in
\eqref{eqn:posterior} by

\begin{align}
  \lb \thetavec | \svec \rb &\approx \int \lb \thetavec | \muvec = \muvec^*\rb
  \lb \muvec^* | \svec \rb d\muvec^*. \label{eqn:mult_impute}
\end{align}

We can evaluate the integral in \eqref{eqn:mult_impute} up to a constant of
proportionality by drawing a realization from $\lb \muvec^*|\svec \rb$ at every
iteration of our MCMC algorithm, and updating model parameters $\thetavec$
conditioned on the realization.

\citet{johnson_continuous-time_2008} introduced a continuous-time correlated
random walk model for movement with measurement error that relies on an
Ornstein-Uhlenbeck process for velocity, and treats the observed paths for each individual as
conditionally independent (i.e., $\lb \svec_i | \muvec^* \rb = \lb \svec_i |
\muvec^*_i \rb$). Continuing with the same model, \citet{johnson_bayesian_2011}
provided an approach for sampling from the posterior predictive path, $\lb
\muvec^*_i | \svec_i \rb$, which we use to evaluate the integral in
(\ref{eqn:mult_impute}).

We approximate the desired posterior using the following two-step procedure:
\begin{enumerate}
\item Draw $K$ different realizations from
  $\lb \muvec^* | \svec \rb$ using the \verb=R= package \verb=crawl=
  (\citealt{johnson_continuous-time_2008}). 
\item At each iteration of the MCMC sampler, draw one of the $K$ samples and
  condition on $\muvec^*$ for parameter updates.
\end{enumerate}
Choosing too small a value for $K$ will result in inference for the
social network that does not properly account for the uncertainty in $\muvec$
arising due to measurement error and temporal asynchronicity, and can
potentially be biased depending on the particular draws from $\lb \muvec^* |
\svec \rb$. In practice, we found a sufficiently large $K$ in our application to
be on the order of 50, as parameter estimates were essentially unchanged for larger $K$. By making use of the two-stage sequential procedure, we are fitting a close approximation to the full Bayesian hierarchical model.

\subsection{Priors}
To demonstrate the value of our model when little is known \textit{a priori}
about the social ties in a study population, we specify diffuse priors for most
parameters in both the simulation and application. We select conjugate
parametric families whenever possible. The priors used in our simulation and
application are shown in the right columns of Tables \ref{tab:simulation} and
\ref{tab:parameters_kw}. While more informative priors could be used when expert
knowledge is available, we found most parameters to be insensitive to the choice
of hyperparameters. The one exception is the network stability parameter $\phi$
(see Section \ref{sec:process_network}). The stability of the network determines the
range of temporal dependence in the dynamic social network. Similar to analogous
range parameters in the geostatistical setting (see, for example, Chapter 6 of
\citealt{Gelfand2010}), $\phi$ can prove difficult to estimate from the data. In
our application (Section \ref{sec:killer_whales}), we used an informative prior that implies
a strongly stable network because we expected the social network to change
slowly relative to the time scale at which the telemetry data were gathered.

\section{Simulation}
\label{sec:simulation}

The primary parameters of scientific interest are in the network $\W$. Thus, we
evaluate the quality of our model by assessing its ability to recover the
network. A baseline model for comparison is one using only proximity as a
criterion for social connectivity. We consider the proximity-based network
defined by

\begin{align}
W^R_{ij}(t) = I_{||\muvec_i - \muvec_j||_2 < R}. \label{eqn:naive}
\end{align}

Though it does not explicitly incorporate the behaviors of attraction and
alignment, defining the network using (\ref{eqn:naive}) is computationally cheap
and closely mirrors the way some data are collected in the field
(\citealt{levin_performance_2015, goldenberg_controlling_2014}). The
proximity-based approach therefore represents a viable alternative against which
we can compare our model. However, failing to consider attraction and alignment
effects, as well as temporal stability in a dynamic social network can lead to
spurious associations that arise when two unconnected individuals happen to pass
each other by chance. Our simulation shows that our model is able to avoid such
pitfalls.

In the following simulation, we generate directly from the proposed process
model and fit the model using paths $\muvec$. We use parameter values (shown in
Table \ref{tab:simulation}) that generate paths closely resembling the data in
our application for killer whales. Details of the methods we used to fit the
model, and the \verb=R= code used to produce this simulation study is provided in Appendix \ref{supp}. We used the posterior mean of $\W$
as a summary of the network, and investigated a variety of radii $R$ with the
proximity-based network, $\W^R$, to define a suite of alternatives. Because we
know the true mean density of the network, $p_1$, we select the proximity-based
network for which the radius yields a mean density as close as possible to the
true value. Choosing a radius that recovers the true mean density would not
generally be possible, thus, we compared our model to a particularly favorable
proximity-based alternative. However, we found that proximity alone provides a
poor estimate of the true network relative to our proposed dynamic network
model.

\begin{figure}[ht]
  \centering
  \includegraphics[width = \linewidth]{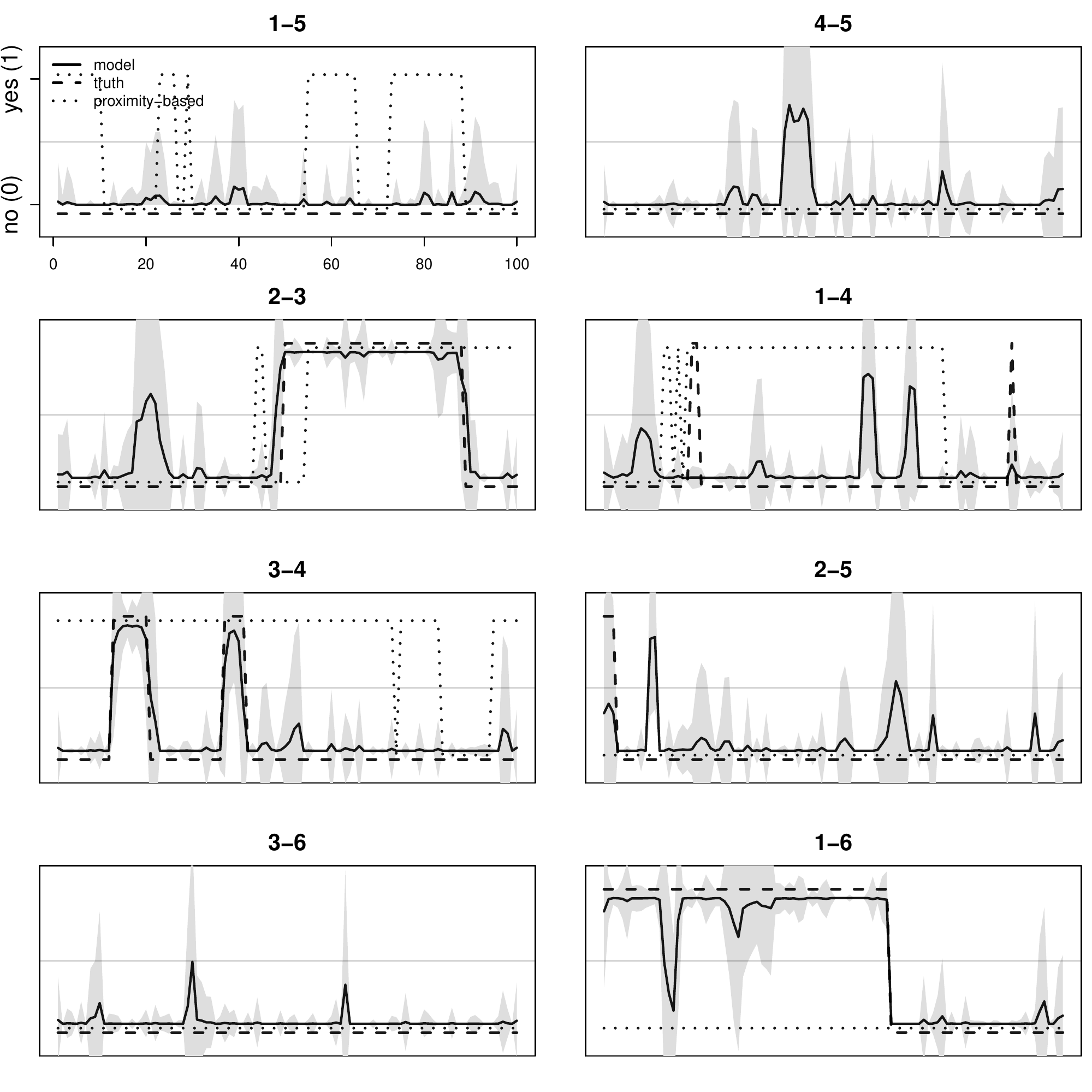}
  \caption{A subset of the complete estimated dynamic network for the simulated
    data on six individuals. The titles correspond to the $i^{th}$ and $j^{th}$
    individuals in $w_{ij}(t)$. The dashed line is the true network, the
    solid line is the posterior mean from the proposed Bayesian model, and
    the gray region represents one standard deviation above and below the
    posterior mean. The dotted line shows the network defined by $\W^R$,
    where individuals are deemed connected whenever they are separated by a
    distance less than $R$ (see Section~\ref{sec:simulation}). (Note: The lines
    are offset slightly near 0 and 1 for visual clarity.)}
  \label{fig:simulation_network}
\end{figure}

Figure \ref{fig:simulation_network} shows estimates of $\W$ for a random
selection of pairs. Included on each plot are the true network (dashed),
the posterior mean from the model fit (solid), and the proximity-based
estimate (dotted). Although the posterior mode of $w_{ij}(t)$ would be a
natural choice for a prediction of the true dynamic social network, we plot the
posterior mean because it provides a visual description about uncertainty in our
predicted network. For example, posterior means of $w_{ij}(t)$ near 0.5 indicate
larger uncertainty about the true connection status of individuals $i$ and $j$
at time $t$ than posterior means near 0 or 1. The pairs 1-5 (top left) and 1-6
(bottom right), show how the proximity-based network can both find spurious
connections, and fail to identify connected behavior when it takes place over
too large a distance. Table \ref{tab:simulation} shows 95\% credible intervals
for all parameters in the model except $\W$. All credible intervals capture the
true parameter values, except those for $\phi$. We observed moderate systematic
bias in the posterior distribution of $\phi$ toward zero, however posterior
inferences for $\W$ were robust despite the bias in $\phi$. In most applications
we expect that the primary questions of scientific interest concern the network
$\W$, and $\phi$ can be treated as a nuisance parameter.

\begin{table}[ht]
  \centering
  \begin{tabular}{r|c|cc|c}
    \multicolumn{1}{c}{} & \multicolumn{1}{c}{} & \multicolumn{2}{c}{posterior} & prior \\
    \hline
    parameter & true & median & (2.5\%, 97.5\%) & density \\
    \hline
     $\alpha$   & 0.9  & 0.92 & (0.77, 0.96) & $\text{Unif}(-1, 1)$        \\
     $\beta$    & 0.5  & 0.46 & (0.37, 0.55) & $\N(0, 10^{3})$             \\
     $p_1$      & 0.2  & 0.15 & (0.0096, 0.21) & $\text{Unif}(0, 1)$         \\
     $\phi$     & 0.95 & 0.84 & (0.78, 0.90) & $\Beta(17.2, 1.5)$ \\
     $c$        & 0.33 & 0.30 & (0.24, 2.36) & $\IG(1.5, 3.5)$             \\
     $\sigma^2$ & 1    & 0.92 & (0.74, 7.91) & $\IG(10^{-1}, 10^{-3})$     \\
     \hline
  \end{tabular}
  \caption{Marginal posterior medians and 95\% credible
    intervals for model parameters. True values for the simulation were chosen
    to yield plausible movement paths. The right column describes
    the prior distributions used.}
  \label{tab:simulation}
\end{table}

Any study of a social network is ultimately based on a definition for connection
specific to the population of interest. Thus, it is incorrect to say that the
proximity-based network fails to capture the true network. Rather, the
proximity-based network simply does a poorer job describing the connections that
influence movement than the network based on our proposed model. It is
impossible to perfectly define a given social network, but if there is reason to
believe that a study population might exhibit the commonly observed behaviors of
attraction and alignment, then our model offers a way to study it. We have shown
that ignoring these mechanisms can result in misleading inference.

\section{Killer whales}
\label{sec:killer_whales}
We analyzed observed data for seven individuals near the Antarctic Peninsula
over the course of a week in February 2013 (for a description of the tags and
study area see \citealt{durban_antarctic_2012,
  andrews_satellite_2008}). Geographic positions were measured using Argos
transmitter tags. Within the study area, three genetically distinct types of
killer whales (termed A, B1, and B2) are known to exist (\citealt{Durban2016, Morin2015,
  pitman_three_2003}) and are characterized primarily by their size, coloration,
and diet. Type A killer whales are the largest and feed primarily on Antarctic
minke whales (\textit{Balaenoptera bonaerensis})
(\citealt{pitman_three_2003}). Of the two type B killer whales, B1 is larger and
is distinguished by a diet consisting primarily of ice seals
(\citealt{durban_antarctic_2012}). Finally, type B2 killer whales are
distinguished by an observed diet of penguins and likely also fish during deep dives
(personal communication J. W. Durban 2015;
\citealt{pitman_killer_2010}). Although all types of killer whales have been
observed exhibiting social behavior within type, association between types has
not been observed. The study sample of seven tagged whales consisted of three
whales of Type A, one of type B1 and three of type B2.

Credible intervals for all parameters except the network $\W$ are shown in Table
\ref{tab:parameters_kw}. When we examine the mean step size across all
individuals and times, we found it to be several times larger than the
contribution of attraction, suggesting only a moderate attractive effect. The
fit also suggests a strong alignment effect evidenced by the posterior median
for $\alpha$ near 1. Therefore, we conclude that connectivity in this population
of killer whales manifests itself predominantly as movement in parallel, with
some additional tendency for connected individuals to move toward one another.

\begin{table}[ht]
  \centering
  \begin{tabular}{r|cc|c}
    \multicolumn{1}{c}{} & \multicolumn{2}{c}{posterior} & prior \\
    \hline
    parameter & median & (2.5\%, 97.5\%) & density \\
    \hline
     $\alpha$ & 0.88   & (0.40,     0.94) & $\text{Unif}(-1, 1)$        \\
     $\beta$  & 0.022  & (0.012,   0.030) & $\N(0, 10^{3})$             \\
     $p_1$    & 0.11   & (0.005,    0.20) & $\text{Unif}(0, 1)$         \\
     $\phi$   & 0.95   & (0.90,     0.98) & $\Beta(100, \frac{100}{9})$ \\
     $c$      & 0.35   & (0.24,     2.87) & $\IG(1.5, 3.5)$             \\
     $\sigma$ & 0.0033 & (0.0026,  0.025) & $\IG(10^{-1}, 10^{-3})$     \\
     \hline
  \end{tabular}
  \caption{Marginal posterior medians and 95\% credible intervals for model
    parameters when fit to the killer whale tagging data. The values reflect a
    strong alignment effect ($\alpha$), weak attraction effect ($\beta$), and a
    stable ($\phi$), sparse ($p_1$) social network. The right column describes
    the prior distributions used.}
  \label{tab:parameters_kw}
\end{table}

\begin{figure}[ht]
  \centering
  \includegraphics[width = \linewidth]{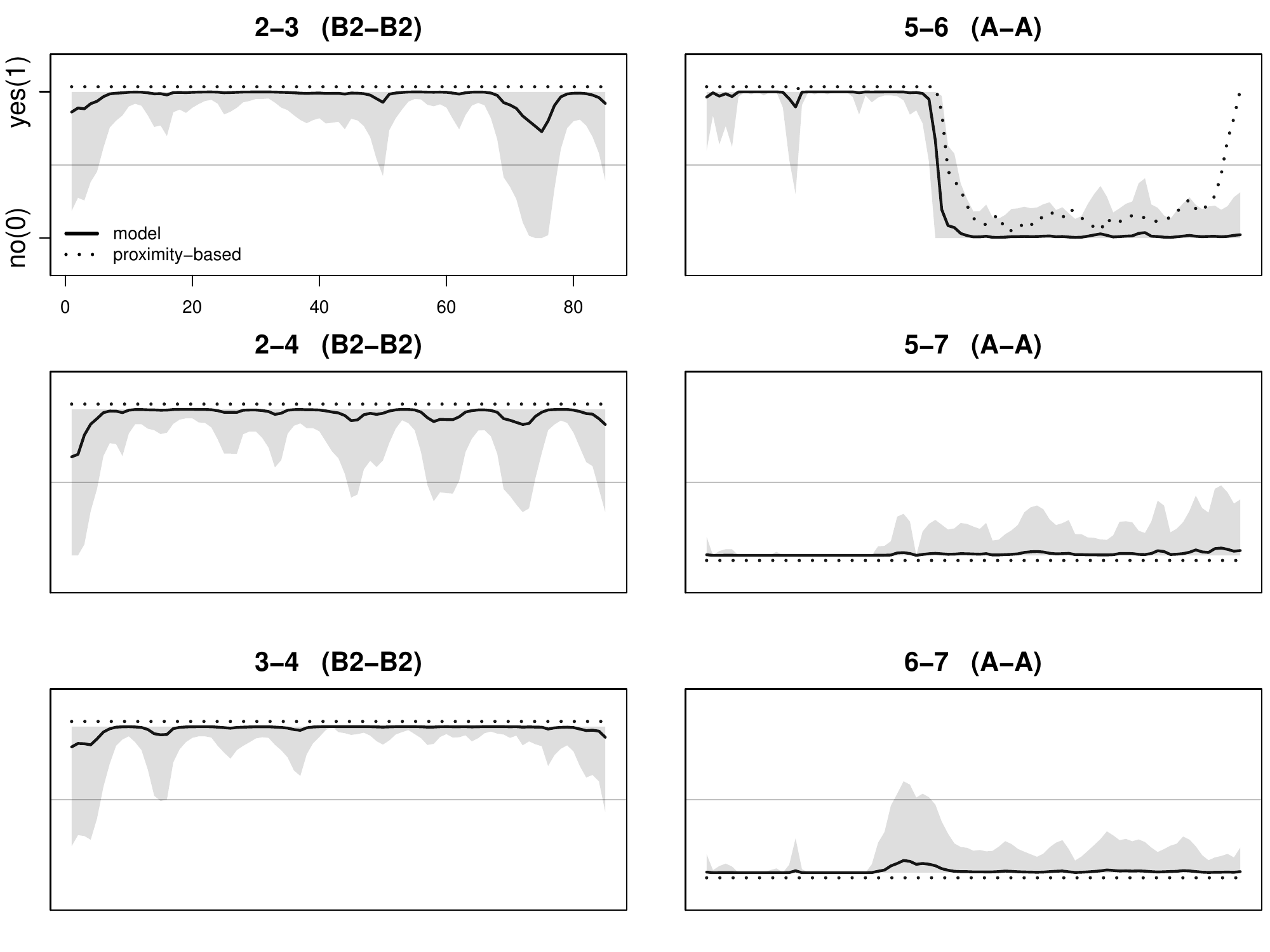}
  \caption{A selection of the ${7 \choose 2} = 21$ possible pairs of individuals
    in the killer whale study sample. The left column is all pairs of killer
    whales of type B2 (labeled 2, 3, 4), and the right column is all pairs of
    killer whales of type A (labeled 5, 6, 7). The solid line in each plot shows
    the posterior mean for $w_{ij}$ and the gray region represents one standard
    deviation above and below the posterior mean. The dotted line shows the
    network defined by $\W^R$, where individuals are deemed connected whenever
    they are separated by a distance less than $R$. (Note: The lines are offset
    slightly near 0 and 1 for visual clarity.)}
  \label{fig:network_kw}
\end{figure}

The credible intervals for $p_1$ and $\phi$ suggest that the network is very
stable, but also fairly sparse. Enduring connections are directly visible in
Figure \ref{fig:network_kw}. The left column shows all pairwise dynamics between
the three individuals of type B2, and the right column shows all pairwise
dynamics between the three individuals of type A. All three individuals of type
B2 show strong connection through the study period and, in fact, all three of
these individuals moved as a group during this time. The only social interaction
involving individuals in type A occurred during the first few days of the study
period between individuals 5 and 6. There was strong evidence for complete
independence between all individuals not in the same type (see Figures
\ref{fig:complete1} and \ref{fig:complete2} in the appendix), consistent with
expert knowledge. Of the 15 inter-type connections in $\W$, there were no
posterior means above 0.5 at any time in the study period. A visualization of
the movement and estimated social connections between these individuals can be
found in Appendix \ref{supp}.

As in the simulation (Section \ref{sec:simulation}), we investigated an
alternative definition for the social network, based purely on proximity, given
by \eqref{eqn:naive}. To account for the uncertainty in $\muvec$, we constructed
the proximity-based network defined by a particular choice of $R$ for each of
the $K$ draws from $\lb \muvec^* | \svec \rb$ used for multiple imputation (see
Section \ref{sec:data_crawl}), and averaged across these networks. The primary
means of communication at a distance between killer whales is acoustic
signaling. Therefore, we selected values for $R$ based on the typical distances
across which killer whales are known to communicate
acoustically. \citet{miller2006} observed killer whales in the Pacific Northwest
and estimated signals between individuals were detectable at distances of 5-15km. This range is consistent with expert knowledge about the killer
whales in our study region. We inspected the corresponding dynamic social
networks for radii between 5-15km and found little variation in the resulting
networks. Figures \ref{fig:network_kw}, \ref{fig:complete1}, and
\ref{fig:complete2} show the proximity-based network for $R$ = 10km.

While we observed some similarities in the proximity-based and model-based
networks, there are several notable discrepancies. For instance, all proximity-based networks for radii between 5-15km included numerous connections between individuals of different types (Figures \ref{fig:complete1} and \ref{fig:complete2}). The presence of inter-type connections conflicts with expert knowledge that killer whales of differing types do not form social bonds, suggesting that the proximity-based network may be defining spurious social ties. Moreover, because the proximity-based network does not account for temporal stability in social connections, we observe instances of implausibly rapid oscillation in connection status (Figure \ref{fig:complete2}). The proximity-based networks and our model-based network provide similar inference for within-type ties (Figure \ref{fig:network_kw}), but our model-based approach also provides rigorous uncertainty estimates. A researcher might arguably make an \textit{ad hoc} adjustment to the network and simply discard all inter-type connections on the basis of prior knowledge, thereby arriving at the same conclusion regardless of which rule was used to define the social network. However, the feasibility of such an approach is unique to this study for two reasons. First, supplementary individual-level information, such as killer whale type, is often unavailable. Second, in many populations, the relationship among covariates and social connections is largely unknown, prohibiting covariate-based pruning of the proximity network.

\section{Discussion}
\label{sec:discussion}
Existing methods for measuring and studying dynamic social networks in animal
populations typically involve \textit{ad hoc} definitions for connectivity based
on direct observation of study populations. Our model offers a flexible, but
interpretable, hierarchical framework that allows researchers to rigorously
study dynamic social networks informed by relatively inexpensive telemetry
data. Moreover, our proposed model can easily be coupled with existing analyses
on dynamic networks. Fundamentally, the study of dynamic social networks often
begins with descriptive statistics such as network density, node degree,
transitivity, and others (\citealt{pinter-wollman_dynamics_2013}). All of these
common summaries can be obtained as derived quantities in our Bayesian framework
with estimates of uncertainty. More sophisticated models for dynamic networks
(e.g., \citealt{durante_nonparametric_2014, sarkar_dynamic_2005,
  sewell_latent_2015}) can take the posterior mode of $\W$ as input, or be
incorporated as part of a larger hierarchical modeling structure.

We have shown, through simulation, that our proposed model is able to capture
information about a population's social structure in a way that a simplistic
proximity-based measure cannot, both by avoiding spurious connections and
detecting interactions that occur over large distances. Through an application
on killer whale movement, we showed that the model captures connections
consistent with expert knowledge based on non-quantitative observation, and can
therefore be relied upon to deliver credible and practical inference.

When auxiliary covariates are available on the individuals, the proposed model
can be extended to include such data. A potential generalization is to allow the
spatial covariates to influence the mean position process of each individual,
$\muvec_i(t)$, linearly. If we denote the matrix containing spatial covariates
$\X_C(t)$, we arrive at a familiar additive form

\begin{align}
  \lb \muvec(t)|\muvec(t - 1), \thetavec, \gammavec\rb &=
  \N \lp \underbrace{\X_C(t - 1)\gammavec}_\text{covariate effect}
  + \underbrace{\X_W(t - 1)\betavec}_\text{attraction}, \;
  \underbrace{\Q(t)}_\text{alignment} \rp
\end{align}

where

\begin{align}
  \X_W(t - 1) \equiv
  \begin{bmatrix} \muvec(t - 1) & \muvectil(t - 1) \end{bmatrix},
  \hspace{0.5in} \betavec \equiv \begin{bmatrix} 1 \\ \beta \end{bmatrix}.
\end{align}

One limitation of our model is that it is computationally intensive for large
study samples. The number of parameters in our model grows at a rate of ${n
  \choose 2} T$ as the number of individuals, $n$, and number of time points,
$T$, increase. The most dominant factor in computation time is typically $n$,
and when the number of individuals grows beyond a few dozen, fitting the model
on a laptop computer using MCMC becomes infeasible. One way to decrease the
computational cost of fitting the model is to introduce additional structure on
$\W$. We suggest two possible approaches.

The first way to introduce structure to $\W$ is to define a maximum radius of
interaction, $R_{\text{max}}$, beyond which the probability of a social
connection is zero. For example, the radius might be chosen to be the maximum
distance at which two individuals are able to detect one another. After
modifying the conditional distribution of $w_{ij}(t)$ based on $R_{\text{max}}$,
it is no longer necessary to update all $w_{ij}(t)$ in each step of the MCMC
algorithm, only those for which $||\muvec_i(t) - \muvec_j(t)|| <
R_{\text{max}}$. If $R_{\text{max}}$ is small relative to the spatial extent of
the trajectories, this proximity-based modification offers a substantial
reduction in the computational cost of fitting the model. This idea is somewhat
related to covariance tapering for spatially referenced Gaussian random
variables. \citet{furrer_covariance_2006} decrease the computational burden of
interpolating, or kriging, by deliberately introducing zeros into the covariance
matrix. In our setting, we would introduce zeros into the precision matrix.

Another way to alleviate the computational burden is to enforce structure
directly on $\W$ to reduce number of parameters in the model. For instance, it
may be reasonable to assume that the social connections in a given population
form as complete subgroups or cliques. In this case, the network describes a
clustering process with only $nT$ parameters. Though motivated by
straightforward mechanisms, both of these approaches to reducing the
computational burden are non-trivial to implement. In the first case, setting a
maximum radius of interaction complicates the enforcement of stability in the
density of the network (introduced in Section \ref{sec:process_network}) and
offers modest or no gains when $R_{\text{max}}$ is large relative to the spatial
extent of the individual paths. In the second case, updating the clustering
process $\W$ requires the exploration of a very large space (of cardinality
equal to the Bell number $B_n$) for every $t$.

Although further developments are required before data for large populations of
individuals can be accommodated, our framework provides a strong foundation for
modeling relationships between movement and social networks.

\section{Acknowledgments}
Killer whale tagging was conducted under permit \#14097 from the National Marine
Fisheries Service and Antarctic Conservation Act permit \#2009-013. Shipboard
tagging operations were supported by Lindblad Expeditions and the National
Geographic Society, and by an NSF rapid grant to Ari Friedlaender. Robert Pitman
helped with tag deployments and identification of killer whale types in the
field.

This material is based in part upon research supported by NOAA ACK
188000 and NSF SES-1461495 and DMS-1614392. Any use of trade,
firm, or product names is for descriptive purposes only and does not imply
endorsement by the U.S. Government.

We would like to thank multiple anonymous reviewers for providing valuable feedback on this manuscript.

\nocite{Scharf2016supp}
\bibliographystyle{imsart-nameyear}
\bibliography{LSSM_manuscript.bib}

\begin{thebibliography}{45}

\bibitem[\protect\citeauthoryear{Andrews, Pitman and
  Ballance}{2008}]{andrews_satellite_2008}
\begin{barticle}[author]
\bauthor{\bsnm{Andrews},~\bfnm{Russel~D.}\binits{R.~D.}},
  \bauthor{\bsnm{Pitman},~\bfnm{Robert~L.}\binits{R.~L.}} \AND
  \bauthor{\bsnm{Ballance},~\bfnm{Lisa~T.}\binits{L.~T.}}
(\byear{2008}).
\btitle{Satellite tracking reveals distinct movement patterns for {type} {B}
  and {type} {C} killer whales in the southern {Ross} {Sea}, {Antarctica}}.
\bjournal{Polar Biology}
\bvolume{31}
\bpages{1461--1468}.
\end{barticle}
\endbibitem

\bibitem[\protect\citeauthoryear{Baird and Whitehead}{2000}]{baird_social_2000}
\begin{barticle}[author]
\bauthor{\bsnm{Baird},~\bfnm{Robin~W}\binits{R.~W.}} \AND
  \bauthor{\bsnm{Whitehead},~\bfnm{Hal}\binits{H.}}
(\byear{2000}).
\btitle{Social organization of mammal-eating killer whales: group stability and
  dispersal patterns}.
\bjournal{Canadian Journal of Zoology}
\bvolume{78}
\bpages{2096--2105}.
\end{barticle}
\endbibitem

\bibitem[\protect\citeauthoryear{Berliner}{1996}]{berliner_hierarchical_1996}
\begin{bincollection}[author]
\bauthor{\bsnm{Berliner},~\bfnm{L.~Mark}\binits{L.~M.}}
(\byear{1996}).
\btitle{Hierarchical {Bayesian} {time} {series} {models}}.
In \bbooktitle{Maximum {Entropy} and {Bayesian} {Methods}},
(\beditor{\bfnm{Kenneth~M.}\binits{K.~M.}~\bsnm{Hanson}} \AND
  \beditor{\bfnm{Richard~N.}\binits{R.~N.}~\bsnm{Silver}}, eds.).
\bseries{Fundamental {Theories} of {Physics}}
\bvolume{79}
\bpages{15--22}.
\bpublisher{Springer Netherlands}.
\end{bincollection}
\endbibitem

\bibitem[\protect\citeauthoryear{Besag}{1974}]{besag_spatial_1974}
\begin{barticle}[author]
\bauthor{\bsnm{Besag},~\bfnm{Julian}\binits{J.}}
(\byear{1974}).
\btitle{Spatial {interaction} and the {statistical} {analysis} of {lattice}
  {systems}}.
\bjournal{Journal of the Royal Statistical Society. Series B (Methodological)}
\bvolume{36}
\bpages{192--236}.
\end{barticle}
\endbibitem

\bibitem[\protect\citeauthoryear{Besag and
  Kooperberg}{1995}]{besag_conditional_1995}
\begin{barticle}[author]
\bauthor{\bsnm{Besag},~\bfnm{Julian}\binits{J.}} \AND
  \bauthor{\bsnm{Kooperberg},~\bfnm{Charles}\binits{C.}}
(\byear{1995}).
\btitle{On {conditional} and {intrinsic} {autoregression}}.
\bjournal{Biometrika}
\bvolume{82}
\bpages{733--746}.
\end{barticle}
\endbibitem

\bibitem[\protect\citeauthoryear{Brillinger and
  Stewart}{1998}]{brillinger_elephant-seal_1998}
\begin{barticle}[author]
\bauthor{\bsnm{Brillinger},~\bfnm{David~R.}\binits{D.~R.}} \AND
  \bauthor{\bsnm{Stewart},~\bfnm{Brent~S.}\binits{B.~S.}}
(\byear{1998}).
\btitle{Elephant-seal movements: {Modelling} migration}.
\bjournal{Canadian Journal of Statistics}
\bvolume{26}
\bpages{431--443}.
\end{barticle}
\endbibitem

\bibitem[\protect\citeauthoryear{Codling and Bode}{2014}]{codling_copycat_2014}
\begin{barticle}[author]
\bauthor{\bsnm{Codling},~\bfnm{Edward~A.}\binits{E.~A.}} \AND
  \bauthor{\bsnm{Bode},~\bfnm{N.~W.}\binits{N.~W.}}
(\byear{2014}).
\btitle{Copycat dynamics in leaderless animal group navigation}.
\bjournal{Movement Ecolcology}
\bvolume{2}
\bpages{11}.
\end{barticle}
\endbibitem

\bibitem[\protect\citeauthoryear{Croft, James and
  Krause}{2008}]{croft_exploring_2008}
\begin{bbook}[author]
\bauthor{\bsnm{Croft},~\bfnm{Darren~P.}\binits{D.~P.}},
  \bauthor{\bsnm{James},~\bfnm{Richard}\binits{R.}} \AND
  \bauthor{\bsnm{Krause},~\bfnm{Jens}\binits{J.}}
(\byear{2008}).
\btitle{Exploring {animal} {social} {networks}}.
\bpublisher{Princeton University Press}.
\end{bbook}
\endbibitem

\bibitem[\protect\citeauthoryear{Durante and
  Dunson}{2014}]{durante_nonparametric_2014}
\begin{barticle}[author]
\bauthor{\bsnm{Durante},~\bfnm{Daniele}\binits{D.}} \AND
  \bauthor{\bsnm{Dunson},~\bfnm{David~B.}\binits{D.~B.}}
(\byear{2014}).
\btitle{Nonparametric {Bayes} dynamic modeling of relational data}.
\bjournal{Biometrika}
\bvolume{101}
\bpages{883--898}.
\end{barticle}
\endbibitem

\bibitem[\protect\citeauthoryear{Durban and
  Pitman}{2012}]{durban_antarctic_2012}
\begin{barticle}[author]
\bauthor{\bsnm{Durban},~\bfnm{J.~W.}\binits{J.~W.}} \AND
  \bauthor{\bsnm{Pitman},~\bfnm{R.~L.}\binits{R.~L.}}
(\byear{2012}).
\btitle{Antarctic killer whales make rapid, round-trip movements to subtropical
  waters: evidence for physiological maintenance migrations?}
\bjournal{Biology Letters}
\bvolume{8}
\bpages{274--277}.
\end{barticle}
\endbibitem

\bibitem[\protect\citeauthoryear{Durban et~al.}{2016}]{Durban2016}
\begin{barticle}[author]
\bauthor{\bsnm{Durban},~\bfnm{J.~W.}\binits{J.~W.}},
  \bauthor{\bsnm{Fearnbach},~\bfnm{H.}\binits{H.}},
  \bauthor{\bsnm{Burrows},~\bfnm{D.~G.}\binits{D.~G.}},
  \bauthor{\bsnm{Ylitalo},~\bfnm{G.~M.}\binits{G.~M.}} \AND
  \bauthor{\bsnm{Pitman},~\bfnm{R.~L.}\binits{R.~L.}}
(\byear{2016}).
\btitle{{Morphological and ecological evidence for two sympatric forms of Type
  B killer whale around the Antarctic Peninsula}}.
\bjournal{Polar Biology}
\bvolume{April}
\bpages{1--6}.
\bdoi{10.1007/s00300-016-1942-x}
\end{barticle}
\endbibitem

\bibitem[\protect\citeauthoryear{Forester
  et~al.}{2007}]{forester_statespace_2007}
\begin{barticle}[author]
\bauthor{\bsnm{Forester},~\bfnm{James~D.}\binits{J.~D.}},
  \bauthor{\bsnm{Ives},~\bfnm{Anthony~R.}\binits{A.~R.}},
  \bauthor{\bsnm{Turner},~\bfnm{Monica~G.}\binits{M.~G.}},
  \bauthor{\bsnm{Anderson},~\bfnm{Dean~P.}\binits{D.~P.}},
  \bauthor{\bsnm{Fortin},~\bfnm{Daniel}\binits{D.}},
  \bauthor{\bsnm{Beyer},~\bfnm{Hawthorne~L.}\binits{H.~L.}},
  \bauthor{\bsnm{Smith},~\bfnm{Douglas~W.}\binits{D.~W.}} \AND
  \bauthor{\bsnm{Boyce},~\bfnm{Mark~S.}\binits{M.~S.}}
(\byear{2007}).
\btitle{State-space models link elk movement patterns to landscape
  characteristics in yellowstone national park}.
\bjournal{Ecological Monographs}
\bvolume{77}
\bpages{285--299}.
\end{barticle}
\endbibitem

\bibitem[\protect\citeauthoryear{Franz
  et~al.}{2015}]{franz_self-organizing_2015}
\begin{barticle}[author]
\bauthor{\bsnm{Franz},~\bfnm{Mathias}\binits{M.}},
  \bauthor{\bsnm{McLean},~\bfnm{Emily}\binits{E.}},
  \bauthor{\bsnm{Tung},~\bfnm{Jenny}\binits{J.}},
  \bauthor{\bsnm{Altmann},~\bfnm{Jeanne}\binits{J.}} \AND
  \bauthor{\bsnm{Alberts},~\bfnm{Susan~C.}\binits{S.~C.}}
(\byear{2015}).
\btitle{Self-organizing dominance hierarchies in a wild primate population}.
\bjournal{Proceedings of the Royal Society B: Biological Sciences}
\bvolume{282}
\bpages{20151512}.
\end{barticle}
\endbibitem

\bibitem[\protect\citeauthoryear{Furrer, Genton and
  Nychka}{2006}]{furrer_covariance_2006}
\begin{barticle}[author]
\bauthor{\bsnm{Furrer},~\bfnm{Reinhard}\binits{R.}},
  \bauthor{\bsnm{Genton},~\bfnm{Marc~G.}\binits{M.~G.}} \AND
  \bauthor{\bsnm{Nychka},~\bfnm{Douglas}\binits{D.}}
(\byear{2006}).
\btitle{Covariance {tapering} for {interpolation} of {large} {spatial}
  {datasets}}.
\bjournal{Journal of Computational and Graphical Statistics}
\bvolume{15}
\bpages{502--523}.
\end{barticle}
\endbibitem

\bibitem[\protect\citeauthoryear{Gelfand et~al.}{2010}]{Gelfand2010}
\begin{bbook}[author]
\bauthor{\bsnm{Gelfand},~\bfnm{Alan~E}\binits{A.~E.}},
  \bauthor{\bsnm{Diggle},~\bfnm{Peter}\binits{P.}},
  \bauthor{\bsnm{Guttorp},~\bfnm{Peter}\binits{P.}} \AND
  \bauthor{\bsnm{Fuentes},~\bfnm{Montserrat}\binits{M.}}
(\byear{2010}).
\btitle{{Handbook of Spatial Statistics}}.
\bdoi{doi:10.1201/9781420072884-c6}
\end{bbook}
\endbibitem

\bibitem[\protect\citeauthoryear{Goldenberg
  et~al.}{2014}]{goldenberg_controlling_2014}
\begin{barticle}[author]
\bauthor{\bsnm{Goldenberg},~\bfnm{Shifra~Z.}\binits{S.~Z.}},
  \bauthor{\bparticle{de} \bsnm{Silva},~\bfnm{Shermin}\binits{S.}},
  \bauthor{\bsnm{Rasmussen},~\bfnm{Henrik~B.}\binits{H.~B.}},
  \bauthor{\bsnm{Douglas-Hamilton},~\bfnm{Iain}\binits{I.}} \AND
  \bauthor{\bsnm{Wittemyer},~\bfnm{George}\binits{G.}}
(\byear{2014}).
\btitle{Controlling for behavioural state reveals social dynamics among male
  {African} elephants, {\textit{Loxodonta}} \textit{africana}}.
\bjournal{Animal Behaviour}
\bvolume{95}
\bpages{111--119}.
\end{barticle}
\endbibitem

\bibitem[\protect\citeauthoryear{Hanks
  et~al.}{2011}]{hanks_velocity-based_2011}
\begin{barticle}[author]
\bauthor{\bsnm{Hanks},~\bfnm{Ephraim~M.}\binits{E.~M.}},
  \bauthor{\bsnm{Hooten},~\bfnm{Mevin~B.}\binits{M.~B.}},
  \bauthor{\bsnm{Johnson},~\bfnm{Devin~S.}\binits{D.~S.}} \AND
  \bauthor{\bsnm{Sterling},~\bfnm{Jeremy~T.}\binits{J.~T.}}
(\byear{2011}).
\btitle{Velocity-{based} {movement} {modeling} for {individual} and
  {population} {level} {inference}}.
\bjournal{PLoS ONE}
\bvolume{6(8)}
\bpages{e22795}.
\end{barticle}
\endbibitem

\bibitem[\protect\citeauthoryear{Hanks et~al.}{2015}]{hanks_restricted_2015}
\begin{barticle}[author]
\bauthor{\bsnm{Hanks},~\bfnm{Ephraim~M.}\binits{E.~M.}},
  \bauthor{\bsnm{Schliep},~\bfnm{Erin~M.}\binits{E.~M.}},
  \bauthor{\bsnm{Hooten},~\bfnm{Mevin~B.}\binits{M.~B.}} \AND
  \bauthor{\bsnm{Hoeting},~\bfnm{Jennifer~A.}\binits{J.~A.}}
(\byear{2015}).
\btitle{Restricted spatial regression in practice: geostatistical models,
  confounding, and robustness under model misspecification}.
\bjournal{Environmetrics}
\bvolume{26}
\bpages{243--254}.
\end{barticle}
\endbibitem

\bibitem[\protect\citeauthoryear{Hooten et~al.}{2010}]{hooten_agent-based_2010}
\begin{barticle}[author]
\bauthor{\bsnm{Hooten},~\bfnm{Mevin~B.}\binits{M.~B.}},
  \bauthor{\bsnm{Johnson},~\bfnm{Devin~S.}\binits{D.~S.}},
  \bauthor{\bsnm{Hanks},~\bfnm{Ephraim~M.}\binits{E.~M.}} \AND
  \bauthor{\bsnm{Lowry},~\bfnm{John~H.}\binits{J.~H.}}
(\byear{2010}).
\btitle{Agent-{based} {inference} for {animal} {movement} and {selection}}.
\bjournal{Journal of Agricultural, Biological, and Environmental Statistics}
\bvolume{15}
\bpages{523--538}.
\end{barticle}
\endbibitem

\bibitem[\protect\citeauthoryear{Johnson, London and
  Kuhn}{2011}]{johnson_bayesian_2011}
\begin{barticle}[author]
\bauthor{\bsnm{Johnson},~\bfnm{Devin~S.}\binits{D.~S.}},
  \bauthor{\bsnm{London},~\bfnm{Josh~M.}\binits{J.~M.}} \AND
  \bauthor{\bsnm{Kuhn},~\bfnm{Carey~E.}\binits{C.~E.}}
(\byear{2011}).
\btitle{Bayesian {inference} for {animal} {space} {use} and {other} {movement}
  {metrics}}.
\bjournal{Journal of Agricultural, Biological, and Environmental Statistics}
\bvolume{16}
\bpages{357--370}.
\end{barticle}
\endbibitem

\bibitem[\protect\citeauthoryear{Johnson
  et~al.}{2008}]{johnson_continuous-time_2008}
\begin{barticle}[author]
\bauthor{\bsnm{Johnson},~\bfnm{Devin~S.}\binits{D.~S.}},
  \bauthor{\bsnm{London},~\bfnm{Joshua~M.}\binits{J.~M.}},
  \bauthor{\bsnm{Lea},~\bfnm{Mary-Anne}\binits{M.-A.}} \AND
  \bauthor{\bsnm{Durban},~\bfnm{John~W.}\binits{J.~W.}}
(\byear{2008}).
\btitle{Continuous-time correlated random walk model for animal telemetry
  data}.
\bjournal{Ecology}
\bvolume{89}
\bpages{1208--1215}.
\end{barticle}
\endbibitem

\bibitem[\protect\citeauthoryear{Jonsen, Flemming and
  Myers}{2005}]{jonsen_robust_2005}
\begin{barticle}[author]
\bauthor{\bsnm{Jonsen},~\bfnm{Ian~D.}\binits{I.~D.}},
  \bauthor{\bsnm{Flemming},~\bfnm{Joanna~Mills}\binits{J.~M.}} \AND
  \bauthor{\bsnm{Myers},~\bfnm{Ransom~A.}\binits{R.~A.}}
(\byear{2005}).
\btitle{Robust state-space modeling of animal movement data}.
\bjournal{Ecology}
\bvolume{86}
\bpages{2874--2880}.
\end{barticle}
\endbibitem

\bibitem[\protect\citeauthoryear{Krause, Croft and
  James}{2007}]{krause_social_2007}
\begin{barticle}[author]
\bauthor{\bsnm{Krause},~\bfnm{J.}\binits{J.}},
  \bauthor{\bsnm{Croft},~\bfnm{D.~P.}\binits{D.~P.}} \AND
  \bauthor{\bsnm{James},~\bfnm{R.}\binits{R.}}
(\byear{2007}).
\btitle{Social network theory in the behavioural sciences: potential
  applications}.
\bjournal{Behavioral Ecology and Sociobiology}
\bvolume{62}
\bpages{15--27}.
\end{barticle}
\endbibitem

\bibitem[\protect\citeauthoryear{Langrock
  et~al.}{2012}]{langrock_flexible_2012}
\begin{barticle}[author]
\bauthor{\bsnm{Langrock},~\bfnm{Roland}\binits{R.}},
  \bauthor{\bsnm{King},~\bfnm{Ruth}\binits{R.}},
  \bauthor{\bsnm{Matthiopoulos},~\bfnm{Jason}\binits{J.}},
  \bauthor{\bsnm{Thomas},~\bfnm{Len}\binits{L.}},
  \bauthor{\bsnm{Fortin},~\bfnm{Daniel}\binits{D.}} \AND
  \bauthor{\bsnm{Morales},~\bfnm{Juan~M.}\binits{J.~M.}}
(\byear{2012}).
\btitle{Flexible and practical modeling of animal telemetry data: hidden
  {Markov} models and extensions}.
\bjournal{Ecology}
\bvolume{93}
\bpages{2336--2342}.
\end{barticle}
\endbibitem

\bibitem[\protect\citeauthoryear{Langrock
  et~al.}{2014}]{langrock_modelling_2014}
\begin{barticle}[author]
\bauthor{\bsnm{Langrock},~\bfnm{Roland}\binits{R.}},
  \bauthor{\bsnm{Hopcraft},~\bfnm{J.~Grant~C.}\binits{J.~G.~C.}},
  \bauthor{\bsnm{Blackwell},~\bfnm{Paul~G.}\binits{P.~G.}},
  \bauthor{\bsnm{Goodall},~\bfnm{Victoria}\binits{V.}},
  \bauthor{\bsnm{King},~\bfnm{Ruth}\binits{R.}},
  \bauthor{\bsnm{Niu},~\bfnm{Mu}\binits{M.}},
  \bauthor{\bsnm{Patterson},~\bfnm{Toby~A.}\binits{T.~A.}},
  \bauthor{\bsnm{Pedersen},~\bfnm{Martin~W.}\binits{M.~W.}},
  \bauthor{\bsnm{Skarin},~\bfnm{Anna}\binits{A.}} \AND
  \bauthor{\bsnm{Schick},~\bfnm{Robert~S.}\binits{R.~S.}}
(\byear{2014}).
\btitle{Modelling group dynamic animal movement}.
\bjournal{Methods in Ecology and Evolution}
\bvolume{5}
\bpages{190--199}.
\end{barticle}
\endbibitem

\bibitem[\protect\citeauthoryear{Lemasson, Anderson and
  Goodwin}{2013}]{lemasson_motion-guided_2013}
\begin{barticle}[author]
\bauthor{\bsnm{Lemasson},~\bfnm{B.~H.}\binits{B.~H.}},
  \bauthor{\bsnm{Anderson},~\bfnm{J.~J.}\binits{J.~J.}} \AND
  \bauthor{\bsnm{Goodwin},~\bfnm{R.~A.}\binits{R.~A.}}
(\byear{2013}).
\btitle{Motion-guided attention promotes adaptive communications during social
  navigation}.
\bjournal{Proceedings of the Royal Society B: Biological Sciences}
\bvolume{280}
\bpages{20122003}.
\end{barticle}
\endbibitem

\bibitem[\protect\citeauthoryear{Levin et~al.}{2015}]{levin_performance_2015}
\begin{barticle}[author]
\bauthor{\bsnm{Levin},~\bfnm{Iris~I.}\binits{I.~I.}},
  \bauthor{\bsnm{Zonana},~\bfnm{David~M.}\binits{D.~M.}},
  \bauthor{\bsnm{Burt},~\bfnm{John~M.}\binits{J.~M.}} \AND
  \bauthor{\bsnm{Safran},~\bfnm{Rebecca~J.}\binits{R.~J.}}
(\byear{2015}).
\btitle{Performance of {Encounternet} {tags}: {field} {tests} of {miniaturized}
  {proximity} {loggers} for {use} on {small} {birds}}.
\bjournal{PLoS ONE}
\bvolume{10}
\bpages{e0137242}.
\end{barticle}
\endbibitem

\bibitem[\protect\citeauthoryear{McClintock
  et~al.}{2014}]{mcclintock_when_2014}
\begin{barticle}[author]
\bauthor{\bsnm{McClintock},~\bfnm{Brett~T.}\binits{B.~T.}},
  \bauthor{\bsnm{Johnson},~\bfnm{Devin~S.}\binits{D.~S.}},
  \bauthor{\bsnm{Hooten},~\bfnm{Mevin~B.}\binits{M.~B.}},
  \bauthor{\bsnm{Hoef},~\bfnm{Jay M.~Ver}\binits{J.~M.~V.}} \AND
  \bauthor{\bsnm{Morales},~\bfnm{Juan~M.}\binits{J.~M.}}
(\byear{2014}).
\btitle{When to be discrete: the importance of time formulation in
  understanding animal movement}.
\bjournal{Movement Ecology}
\bvolume{2}
\bpages{21}.
\end{barticle}
\endbibitem

\bibitem[\protect\citeauthoryear{Miller}{2006}]{miller2006}
\begin{barticle}[author]
\bauthor{\bsnm{Miller},~\bfnm{Patrick J.~O.}\binits{P.~J.~O.}}
(\byear{2006}).
\btitle{{Diversity in sound pressure levels and active space of killer whale
  vocalizations}}.
\bjournal{Journal of Comparative Physiology A}
\bvolume{192}
\bpages{449--459}.
\end{barticle}
\endbibitem

\bibitem[\protect\citeauthoryear{Morales et~al.}{2010}]{morales_building_2010}
\begin{barticle}[author]
\bauthor{\bsnm{Morales},~\bfnm{J.~M.}\binits{J.~M.}},
  \bauthor{\bsnm{Moorcroft},~\bfnm{P.~R.}\binits{P.~R.}},
  \bauthor{\bsnm{Matthiopoulos},~\bfnm{J.}\binits{J.}},
  \bauthor{\bsnm{Frair},~\bfnm{J.~L.}\binits{J.~L.}},
  \bauthor{\bsnm{Kie},~\bfnm{J.~G.}\binits{J.~G.}},
  \bauthor{\bsnm{Powell},~\bfnm{R.~A.}\binits{R.~A.}},
  \bauthor{\bsnm{Merrill},~\bfnm{E.~H.}\binits{E.~H.}} \AND
  \bauthor{\bsnm{Haydon},~\bfnm{D.~T.}\binits{D.~T.}}
(\byear{2010}).
\btitle{Building the bridge between animal movement and population dynamics}.
\bjournal{Philosophical Transactions of the Royal Society B: Biological
  Sciences}
\bvolume{365}
\bpages{2289--2301}.
\end{barticle}
\endbibitem

\bibitem[\protect\citeauthoryear{Morin et~al.}{2015}]{Morin2015}
\begin{barticle}[author]
\bauthor{\bsnm{Morin},~\bfnm{Phillip~A}\binits{P.~A.}},
  \bauthor{\bsnm{Parsons},~\bfnm{Kim~M}\binits{K.~M.}},
  \bauthor{\bsnm{Archer},~\bfnm{Frederick~I}\binits{F.~I.}},
  \bauthor{\bsnm{{\'{A}}vila-Arcos},~\bfnm{Mar{\'{\i}}a~C}\binits{M.~C.}},
  \bauthor{\bsnm{Barrett-Lennard},~\bfnm{Lance~G}\binits{L.~G.}},
  \bauthor{\bsnm{{Dalla Rosa}},~\bfnm{Luciano}\binits{L.}},
  \bauthor{\bsnm{Duch{\^{e}}ne},~\bfnm{Sebasti{\'{a}}n}\binits{S.}},
  \bauthor{\bsnm{Durban},~\bfnm{John~W}\binits{J.~W.}},
  \bauthor{\bsnm{Ellis},~\bfnm{Graeme~M}\binits{G.~M.}},
  \bauthor{\bsnm{Ferguson},~\bfnm{Steven~H}\binits{S.~H.}},
  \bauthor{\bsnm{Ford},~\bfnm{John~K}\binits{J.~K.}},
  \bauthor{\bsnm{Ford},~\bfnm{Michael~J}\binits{M.~J.}},
  \bauthor{\bsnm{Garilao},~\bfnm{Cristina}\binits{C.}},
  \bauthor{\bsnm{Gilbert},~\bfnm{M~Thomas~P}\binits{M.~T.~P.}},
  \bauthor{\bsnm{Kaschner},~\bfnm{Kristin}\binits{K.}},
  \bauthor{\bsnm{Matkin},~\bfnm{Craig~O}\binits{C.~O.}},
  \bauthor{\bsnm{Petersen},~\bfnm{Stephen~D}\binits{S.~D.}},
  \bauthor{\bsnm{Robertson},~\bfnm{Kelly~M}\binits{K.~M.}},
  \bauthor{\bsnm{Visser},~\bfnm{Ingrid~N}\binits{I.~N.}},
  \bauthor{\bsnm{Wade},~\bfnm{Paul~R}\binits{P.~R.}},
  \bauthor{\bsnm{Ho},~\bfnm{Simon Y~W}\binits{S.~Y.~W.}} \AND
  \bauthor{\bsnm{Foote},~\bfnm{Andrew~D}\binits{A.~D.}}
(\byear{2015}).
\btitle{{Geographical and temporal dynamics of a global radiation and
  diversification in the killer whale.}}
\bjournal{Molecular ecology}
\bvolume{24}
\bpages{3964--3979}.
\end{barticle}
\endbibitem

\bibitem[\protect\citeauthoryear{Parsons et~al.}{2009}]{Parsons2009}
\begin{barticle}[author]
\bauthor{\bsnm{Parsons},~\bfnm{K.~M.}\binits{K.~M.}},
  \bauthor{\bsnm{Balcomb},~\bfnm{K.~C.}\binits{K.~C.}},
  \bauthor{\bsnm{Ford},~\bfnm{J.~K.~B.}\binits{J.~K.~B.}} \AND
  \bauthor{\bsnm{Durban},~\bfnm{J.~W.}\binits{J.~W.}}
(\byear{2009}).
\btitle{{The social dynamics of southern resident killer whales and
  conservation implications for this endangered population}}.
\bjournal{Animal Behaviour}
\bvolume{77}
\bpages{963--971}.
\end{barticle}
\endbibitem

\bibitem[\protect\citeauthoryear{Pinter-Wollman
  et~al.}{2013}]{pinter-wollman_dynamics_2013}
\begin{barticle}[author]
\bauthor{\bsnm{Pinter-Wollman},~\bfnm{Noa}\binits{N.}},
  \bauthor{\bsnm{Hobson},~\bfnm{Elizabeth~A.}\binits{E.~A.}},
  \bauthor{\bsnm{Smith},~\bfnm{Jennifer~E.}\binits{J.~E.}},
  \bauthor{\bsnm{Edelman},~\bfnm{Andrew~J.}\binits{A.~J.}},
  \bauthor{\bsnm{Shizuka},~\bfnm{Daizaburo}\binits{D.}},
  \bauthor{\bsnm{Silva},~\bfnm{Shermin~de}\binits{S.~d.}},
  \bauthor{\bsnm{Waters},~\bfnm{James~S.}\binits{J.~S.}},
  \bauthor{\bsnm{Prager},~\bfnm{Steven~D.}\binits{S.~D.}},
  \bauthor{\bsnm{Sasaki},~\bfnm{Takao}\binits{T.}},
  \bauthor{\bsnm{Wittemyer},~\bfnm{George}\binits{G.}},
  \bauthor{\bsnm{Fewell},~\bfnm{Jennifer}\binits{J.}} \AND
  \bauthor{\bsnm{McDonald},~\bfnm{David~B.}\binits{D.~B.}}
(\byear{2013}).
\btitle{The dynamics of animal social networks: analytical, conceptual, and
  theoretical advances}.
\bjournal{Behavioral Ecology}
\bpages{art047}.
\end{barticle}
\endbibitem

\bibitem[\protect\citeauthoryear{Pitman and Durban}{2010}]{pitman_killer_2010}
\begin{barticle}[author]
\bauthor{\bsnm{Pitman},~\bfnm{Robert~L.}\binits{R.~L.}} \AND
  \bauthor{\bsnm{Durban},~\bfnm{John~W.}\binits{J.~W.}}
(\byear{2010}).
\btitle{Killer whale predation on penguins in {Antarctica}}.
\bjournal{Polar Biology}
\bvolume{33}
\bpages{1589--1594}.
\end{barticle}
\endbibitem

\bibitem[\protect\citeauthoryear{Pitman and
  Durban}{2012}]{pitman_cooperative_2012}
\begin{barticle}[author]
\bauthor{\bsnm{Pitman},~\bfnm{Robert~L.}\binits{R.~L.}} \AND
  \bauthor{\bsnm{Durban},~\bfnm{John~W.}\binits{J.~W.}}
(\byear{2012}).
\btitle{Cooperative hunting behavior, prey selectivity and prey handling by
  pack ice killer whales (\textit{{Orcinus} orca}), type {B}, in {Antarctic}
  {Peninsula} waters}.
\bjournal{Marine Mammal Science}
\bvolume{28}
\bpages{16--36}.
\end{barticle}
\endbibitem

\bibitem[\protect\citeauthoryear{Pitman and Ensor}{2003}]{pitman_three_2003}
\begin{barticle}[author]
\bauthor{\bsnm{Pitman},~\bfnm{Robert~L.}\binits{R.~L.}} \AND
  \bauthor{\bsnm{Ensor},~\bfnm{Paul}\binits{P.}}
(\byear{2003}).
\btitle{Three forms of killer whales (\textit{Orcinus orca}) in {Antarctic}
  waters}.
\bjournal{Journal of Cetacean Research and Management}
\bvolume{5}
\bpages{131--140}.
\end{barticle}
\endbibitem

\bibitem[\protect\citeauthoryear{Rue and Held}{2005}]{rue_gaussian_2005}
\begin{bbook}[author]
\bauthor{\bsnm{Rue},~\bfnm{Havard}\binits{H.}} \AND
  \bauthor{\bsnm{Held},~\bfnm{Leonhard}\binits{L.}}
(\byear{2005}).
\btitle{Gaussian {Markov} {random} {fields}: {theory} and {applications}}.
\bpublisher{CRC Press}.
\end{bbook}
\endbibitem

\bibitem[\protect\citeauthoryear{Russell, Hanks and
  Haran}{2015}]{russell_dynamic_2015}
\begin{barticle}[author]
\bauthor{\bsnm{Russell},~\bfnm{James~C.}\binits{J.~C.}},
  \bauthor{\bsnm{Hanks},~\bfnm{Ephraim~M.}\binits{E.~M.}} \AND
  \bauthor{\bsnm{Haran},~\bfnm{Murali}\binits{M.}}
(\byear{2015}).
\btitle{Dynamic models of animal movement with spatial point process
  interactions}.
\bjournal{Journal of Agricultural, Biological, and Environmental Statistics}
\bpages{1--19}.
\end{barticle}
\endbibitem

\bibitem[\protect\citeauthoryear{Sarkar and Moore}{2005}]{sarkar_dynamic_2005}
\begin{barticle}[author]
\bauthor{\bsnm{Sarkar},~\bfnm{Purnamrita}\binits{P.}} \AND
  \bauthor{\bsnm{Moore},~\bfnm{Andrew~W.}\binits{A.~W.}}
(\byear{2005}).
\btitle{Dynamic {social} {network} {analysis} {using} {latent} {space}
  {models}}.
\bjournal{SIGKDD Explor. Newsl.}
\bvolume{7}
\bpages{31--40}.
\end{barticle}
\endbibitem

\bibitem[\protect\citeauthoryear{Scharf et~al.}{2016}]{Scharf2016supp}
\begin{barticle}[author]
\bauthor{\bsnm{Scharf},~\bfnm{Henry}\binits{H.}},
  \bauthor{\bsnm{Hooten},~\bfnm{Mevin~B.}\binits{M.~B.}},
  \bauthor{\bsnm{Fosdick},~\bfnm{Bailey}\binits{B.}},
  \bauthor{\bsnm{Johnson},~\bfnm{Devin}\binits{D.}},
  \bauthor{\bsnm{London},~\bfnm{Josh}\binits{J.}} \AND
  \bauthor{\bsnm{Durban},~\bfnm{John}\binits{J.}}
(\byear{2016}).
\btitle{Supplement to ``Dynamic social networks based on movement"}.
\end{barticle}
\endbibitem

\bibitem[\protect\citeauthoryear{Sewell and Chen}{2015}]{sewell_latent_2015}
\begin{barticle}[author]
\bauthor{\bsnm{Sewell},~\bfnm{Daniel~K}\binits{D.~K.}} \AND
  \bauthor{\bsnm{Chen},~\bfnm{Yuguo}\binits{Y.}}
(\byear{2015}).
\btitle{Latent space models for dynamic networks}.
\bjournal{Journal of the American Statistical Association}
\bvolume{110}
\bpages{1646--1657}.
\end{barticle}
\endbibitem

\bibitem[\protect\citeauthoryear{Sih, Hanser and
  McHugh}{2009}]{sih_social_2009}
\begin{barticle}[author]
\bauthor{\bsnm{Sih},~\bfnm{Andrew}\binits{A.}},
  \bauthor{\bsnm{Hanser},~\bfnm{Sean~F.}\binits{S.~F.}} \AND
  \bauthor{\bsnm{McHugh},~\bfnm{Katherine~A.}\binits{K.~A.}}
(\byear{2009}).
\btitle{Social network theory: new insights and issues for behavioral
  ecologists}.
\bjournal{Behavioral Ecology and Sociobiology}
\bvolume{63}
\bpages{975--988}.
\end{barticle}
\endbibitem

\bibitem[\protect\citeauthoryear{Wey et~al.}{2008}]{wey_social_2008}
\begin{barticle}[author]
\bauthor{\bsnm{Wey},~\bfnm{Tina}\binits{T.}},
  \bauthor{\bsnm{Blumstein},~\bfnm{Daniel~T.}\binits{D.~T.}},
  \bauthor{\bsnm{Shen},~\bfnm{Weiwei}\binits{W.}} \AND
  \bauthor{\bsnm{Jordán},~\bfnm{Ferenc}\binits{F.}}
(\byear{2008}).
\btitle{Social network analysis of animal behaviour: a promising tool for the
  study of sociality}.
\bjournal{Animal Behaviour}
\bvolume{75}
\bpages{333--344}.
\end{barticle}
\endbibitem

\bibitem[\protect\citeauthoryear{Williams and
  Lusseau}{2006}]{williams_killer_2006}
\begin{barticle}[author]
\bauthor{\bsnm{Williams},~\bfnm{Rob}\binits{R.}} \AND
  \bauthor{\bsnm{Lusseau},~\bfnm{David}\binits{D.}}
(\byear{2006}).
\btitle{A killer whale social network is vulnerable to targeted removals}.
\bjournal{Biology Letters}
\bvolume{2}
\bpages{497--500}.
\end{barticle}
\endbibitem

\bibitem[\protect\citeauthoryear{Williams, Trites and
  Bain}{2002}]{williams_behavioural_2002}
\begin{barticle}[author]
\bauthor{\bsnm{Williams},~\bfnm{Rob}\binits{R.}},
  \bauthor{\bsnm{Trites},~\bfnm{Andrew~W.}\binits{A.~W.}} \AND
  \bauthor{\bsnm{Bain},~\bfnm{David~E.}\binits{D.~E.}}
(\byear{2002}).
\btitle{Behavioural responses of killer whales ({Orcinus} orca) to
  whale-watching boats: opportunistic observations and experimental
  approaches}.
\bjournal{Journal of Zoology}
\bvolume{256}
\bpages{255--270}.
\end{barticle}
\endbibitem

\end{thebibliography}

\appendix

\section{Supplemental Materials}
\label{supp}
 \textbf{Supplements A, B, \& C}
 (A) Priors and full-conditionals for the model are presented; (B) Animation of killer whales; (C) Code used for simulation. All can be found here: \url{http://www.stat.colostate.edu/~scharfh/supplemental/}

\section{Dynamic Social Network between killer whale types}
\begin{figure}[h!]
  \centering
  \includegraphics[width = \linewidth]{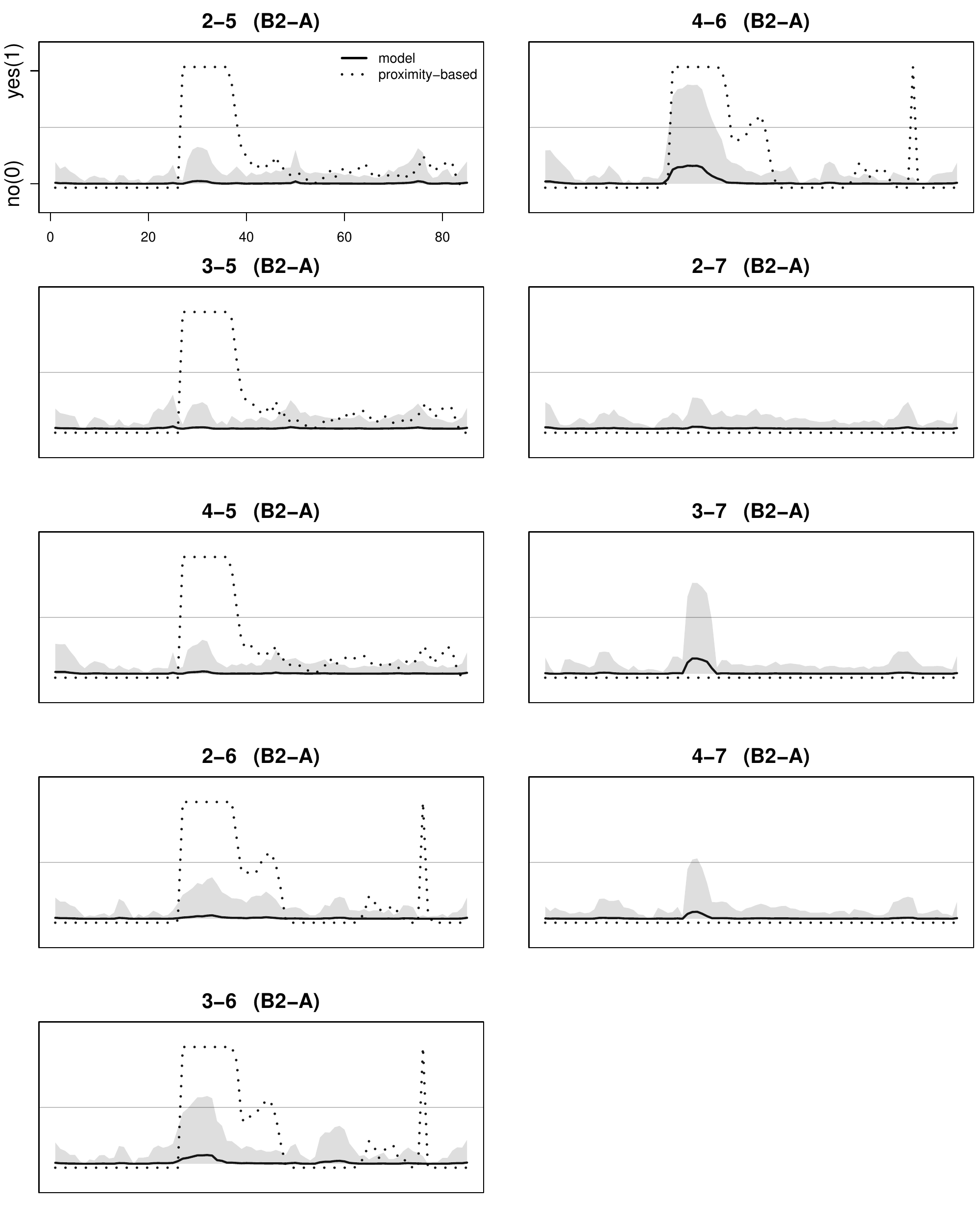}
  \caption{A selection of the ${7 \choose 2} = 21$ possible pairs of individuals
    in the killer whale study sample. The plots displayed are for all inter-type
    pairs of killer whales of type B2 (labeled 2, 3, 4) and A (labeled 5, 6,
    7). The solid line in each plot shows the posterior mean for $w_{ij}$ and
    the gray region represents one standard deviation above and below the
    posterior mean. The dotted line shows the network defined by $\W^R$, where
    individuals are deemed connected whenever they are separated by a distance
    less than $R$. No posterior means above 0.5 were predicted for inter-type
    connections. (Note: The lines are offset slightly near 0 and 1 for visual
    clarity.)}
  \label{fig:complete1}
\end{figure}

\begin{figure}[h!]
  \centering
  \includegraphics[width =
  \linewidth]{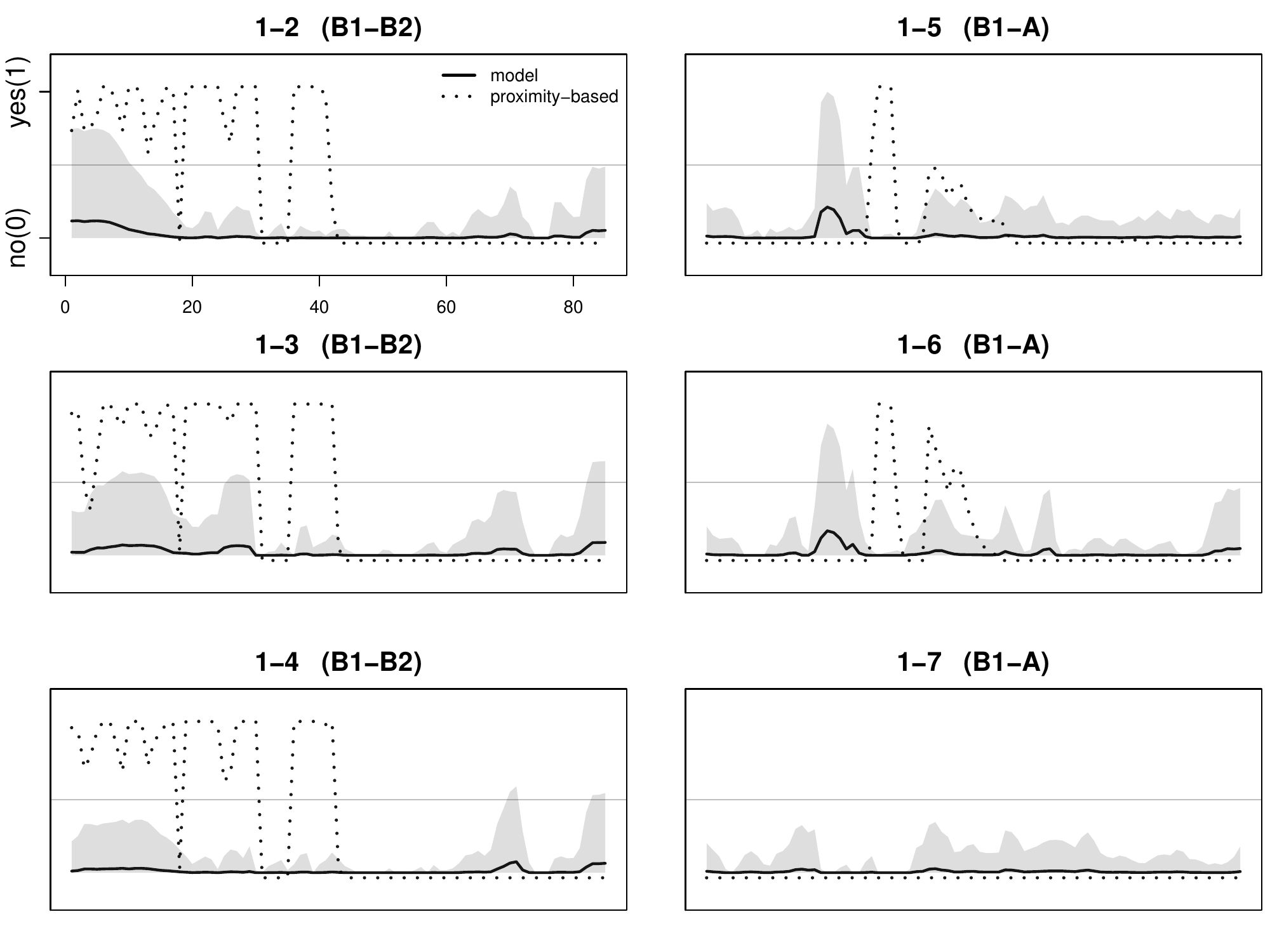}
  \caption{A selection of the ${7 \choose 2} = 21$ possible pairs of individuals
    in the killer whale study sample. The plots displayed are for all inter-type
    pairs of killer whales between the sole individual of type B1 (labeled 1)
    and those of type B2 (labeled 2, 3, 4) and A (labeled 5, 6, 7). The solid
    line in each plot shows the posterior mean for $w_{ij}$ and the gray region
    represents one standard deviation above and below the posterior mean. The
    dotted line shows the network defined by $\W^R$, where individuals are
    deemed connected whenever they are separated by a distance less than $R$. No
    posterior means above 0.5 were predicted for inter-type connections. (Note:
    The lines are offset slightly near 0 and 1 for visual clarity.)}
  \label{fig:complete2}
\end{figure}

\end{document}